\documentclass{jnmp}

\usepackage{amsmath}
\JNMPnumberwithin{equation}{section}

\begin{document}

\renewcommand{\evenhead}{J~Lenells}
\renewcommand{\oddhead}{The Scattering Approach for the Camassa--Holm Equation}

\thispagestyle{empty}

\FirstPageHead{9}{4}{2002}{\pageref{lenells-firstpage}--\pageref{lenells-lastpage}}{Letter}

\copyrightnote{2002}{J~Lenells}

\Name{The Scattering Approach\\
 for the Camassa--Holm equation}
\label{lenells-firstpage}

\Author{Jonatan LENELLS}

\Address{Department of Mathematics, Lund University, P.O.
Box 118, SE-221 00 Lund, Sweden \\
E-mail: jonatan@maths.lth.se}

\Date{Received March 8, 2002; Revised April 23, 2002; Accepted
April 24, 2002}

\begin{abstract}
\noindent We present an approach proving the integrability of the Camassa--Holm
equation for initial data of small amplitude.
\end{abstract}

\section{Introduction}

The Camassa--Holm equation
\begin{equation}\label{ch}
  u_t-u_{txx}+2\omega u_x+3uu_x=2u_xu_{xx}+uu_{xxx},\qquad t>0, \quad  x\in
 {\mathbb R},
\end{equation}
in dimensionless space-time variables $(x,t)$ is a model for the
unidirectional propagation of two-dimensional waves in shallow
water with a flat bottom, $\omega$ being a positive constant
related to the critical shallow water speed (see~\cite{C-H}), and
was first derived as an abstract equation in~\cite{F-F} by using
the method of recursion operators. Of physical interest are
solutions of (\ref{ch}) which decay at infinity cf.~\cite{J}. The
Camassa--Holm equation models wave breaking~\cite{CE} and admits
wave solutions that exist indefinitely in time~\cite{C1}. An
aspect of considerable interest is the fact that the solitary
waves of~(\ref{ch}) are solitons (see~\cite{CHH} for numerical
evidence and~\cite{J2} for the complete description).

In terms of the momentum $m=u-u_{xx}$ the Camassa--Holm
equation can be expressed
cf.~\cite{CHH} as the condition of compatibility between
\begin{equation}\label{s}
  \psi_{xx}=\frac{1}{4}\psi + \lambda (m+ \omega) \psi
\end{equation}
and
\begin{equation}\label{s2}
  \psi_t=\left(\frac{1}{2\lambda}-u\right)\psi_x + \frac{1}{2}u_x\psi.
\end{equation}
Equation (\ref{s}) is the isospectral problem associated to (\ref{ch}) so
that the Camassa--Holm equation is formally integrable. In the absence
of bound states for~(\ref{s}), the direct and
inverse scattering problem was discussed in~\cite{C2}. Our purpose is
to indicate how the scattering approach can be pursued in the more
general case when finitely many bound states are present. In
particular, this allows us to solve the Camassa--Holm initial-value
problem via inverse scattering for initial data of small amplitude.
The importance of the scattering problem in the case of finitely many
bound states is emphasized by the fact that in the
case of one soliton there is precisely one bound state for (\ref{s})
cf.~\cite{CS} and the $2$-soliton solution corresponds to an
isospectral problem (\ref{s}) with two bound states (see \cite {J2}).

\section{Direct and inverse scattering}

If the momentum $m\in H^3({\mathbb R})$ is such that $m+\omega > 0$ and
\[
 \int_{\mathbb R} (1+|x|)|m(x)|dx < \infty,
\]
then the continuous spectrum of (\ref{s}) is $(-\infty, -
\frac{1}{4\omega }]$
and there are at most finitely many eigenvalues in the interval
 $(-\frac{1}{4\omega },0)$ cf.~\cite{C2}. In absence of bound states
 the scattering data consists of the transmission and reflection
 coefficients associated to the elements in the continuous
 spectrum. More precisely, if $\psi (x,t)$ is an eigenfunction
 corresponding to some $\lambda $ in the continuous spectrum of the
 isospectral problem (\ref{s}), then
\begin{equation}
\psi(t,x) \approx \left\{
\begin{array}{ll}
e^{-ikx}+{\mathfrak R}(t,k)\, e^{ikx}\quad & \mbox{as}\quad
 x \to \infty,\vspace{2mm}\\
{\mathfrak T}(t,k)\, e^{-ikx}\quad & \hbox{as}\quad x \to -\infty,
\end{array}\right.
\end{equation}
for some complex transmission coefficient ${\mathfrak T}$ and a reflection
coefficient ${\mathfrak R}$, where $k \ge 0$ satisfies $k^2=-\frac{1}{4}-
\lambda\omega \ge 0$. The evolution of ${\mathfrak T}(t,k)$ and ${\mathfrak
  R}(t,k)$ under the Camassa--Holm flow is given by (see~\cite{C2})
\begin{equation}
{\mathfrak T}(t,k)={\mathfrak T}(0,k), \qquad {\mathfrak R}(t,k)=
{\mathfrak R}(0,k) \exp\left(\frac{ik}{\lambda} \, t\right),\qquad t \ge 0.
\end{equation}
Since (\ref{s}) is the isospectral problem,
the bound states are constants of motion for the
Camassa--Holm equation~\cite{C-H}. In order to solve the scattering problem in
presence of finitely many bound states it is necessary to find the
proper normalization constants for the eigenfunctions associated with
the discrete spectrum. This question was left open in~\cite{C2}, due
to the fact that the choice suggested by analogy with the classical
Schr\"odinger equation~\cite{D-J} is not appropriate (the time
evolution cannot be determined). A proper family of
normalization constants can be defined as follows. The Liouville transformation
\[
\phi(y)= \left(m(x)+\omega\right)^{1/4} \psi(x),
\]
where
\[
y=\int_{0}^x \sqrt{m(\xi)+ \omega}\, d\xi-\int_0^\infty
\frac{m(\xi )}{\sqrt{\omega }+\sqrt{m(\xi )+\omega}} \,d\xi
\]
converts (\ref{s}) into
\begin{equation}
-\frac{d^2 \phi}{d y^2}+ Q \phi=\mu \phi.
\end{equation}
Here
\[
Q(y)=\frac{1}{4q(y)}+\frac{q_{yy}(y)}{4\,q(y)}-
\frac{3 q_y^2(y)}{16 q^2(y)}
 - \frac{1}{4 \omega}\qquad\mbox{with}\quad q(y)=m(x)+\omega,
\]
and the spectral parameter is $\mu=-\frac{1}{4 \omega}-\lambda$. If
$\psi_n(x,t)$ is an eigenfunction for (1.2) corresponding to the
eigenvalue $\lambda_n \in (-\frac{1}{4\omega },0)$, then
\[
\phi_n(y,t)= \left(m(x,t)+\omega\right)^{1/4} \psi_n(x,t)
\]
is an eigenfunction for (2.3) corresponding to the eigenvalue $\mu_n =
-\frac{1}{4\omega }-\lambda_n <0$. Requiring
\[
 \int_{\mathbb R} \phi_n^2 (y,t)\,dy=1,
\]
the normalization constants $c_n(t)$ are determined by
\[
\phi_n (y,t) \approx c_n(t) e^{-k_n y} \qquad \mbox{for}   \quad y
\to \infty,
\]
with $k_n = \sqrt{-\mu_n}$. It turns out that as $m(x,t)$ evolves
according to the Camassa--Holm equation,
\[
c_n(t)=c_n(0) \exp\left(-\frac{k_n \sqrt{\omega }}{2\lambda_n}\, t\right), \qquad
  t \ge 0,
\]
as a rather intricate analysis shows. Therefore the evolution of the
scattering data (the reflection and transmission coefficients together
 with the previously defined normalization constants)
under the Camassa--Holm flow has been explicitly determined. At this point
the method presented in~\cite{C2}
can be implemented to solve the inverse scattering problem
for the Camassa--Holm equation.

 The presented approach is best exemplified by the fact that it shows
 that the solitary waves of~(1.1) are solitons: the associated
 spectral problem is reflectionless and has a single
 eigenvalue. This important feature of the Camassa--Holm equation was
 explained in~\cite{CS} by means of trace formulas
 and eigenvalue estimates for Schr\"odinger operators. An application
 of our technique provides a simpler and more direct proof.

It is known (see~ \cite{LO} and~\cite{CS}) that solitary wave solutions
$u(x,t)=\varphi (x-ct)$ propagating at the speed $c>0$ exist only for
$c>2\omega $. Moreover, the function $\varphi $, determined uniquely
up to translations (henceforth we choose $\varphi $ with the crest
positioned at $x=0$), is smooth and positive with a profile decreasing
symmetrically from its crest of height $(c-2\omega )$. No mathematical
expression in closed form is available for $\varphi $ so that our
analysis depends on the equations
\begin{equation}
-c\varphi + c\varphi_{xx} + \frac{3}{2}\, \varphi^2 +2\omega \varphi =
\varphi \varphi_{xx} + \frac{1}{2}\, \varphi_x^2,
\end{equation}
and
\begin{equation}
\varphi_x^2(c-\varphi ) = \varphi^2 (c-2\omega - \varphi),
\end{equation}
which are both obtained from (1.1). For $u(x,0)=\varphi (x)$ we have
\[
m(x,0)+\omega = \frac{\omega c^2}{[c-\varphi(x)]^2}
\]
so that the Liouville transformation can be performed. In combination
with (2.4)--(2.5), it yields
\[
 Q(y)= -\frac{\varphi (x)}{c[c-\varphi (x)]}.
\]
Straightforward (but long) calculations relying repeatedly on
(2.4)--(2.5) show that $Q(y)$ satisfies the differential equation
\[
Q_y^2 = Q^2\left(2Q+ \frac{c-2\omega }{c\omega }\right).
\]
Hence (see \cite{D-J})
\[
 Q(y) = -\frac{c-2\omega }{2c\omega }\,{\mbox{sech}}^2\left(\frac{1}{2}
\sqrt{\frac{c-2\omega }{c\omega }}\,y\right).
\]
With the above potential it is well-known that the problem (2.3) is
reflectionless and has $\mu=\frac{2\omega-c}{4 c\omega }$
as the only eigenvalue cf.~\cite{D-J}. While performing the Liouville
transformation, let us note that
\begin{equation}
\left\{ \begin{array}{ll}
y(x)-\sqrt{\omega}\,x \to 0\quad& \mbox{as}\quad
 x \to \infty,\vspace{2mm}\\
\displaystyle y(x)-\sqrt{\omega}\,x \to \int_{-\infty}^\infty \left(
\sqrt{\omega}-\sqrt{m(\xi)+\omega}\right) d\xi\quad & \mbox{as}\quad
 x \to -\infty.
\end{array}\right.
\end{equation}

From (1.1) we infer that  $C=\int_{-\infty}^\infty \left(
\sqrt{\omega}-\sqrt{m(\xi)+\omega}\right)d\xi$ is preserved under the
Camassa--Holm flow. Therefore (2.6) can be used to deduce that
\begin{equation}
\omega^{-1/4}\phi(t,y) \approx
\left\{\begin{array}{ll}
e^{-i\sqrt{\mu}y}+{\mathfrak R}(t,k)\, e^{i\sqrt{\mu}y}\quad& \mbox{as}\quad
 y \to \infty,\vspace{2mm}\\
{\mathfrak T}(t,k)\, e^{-i\sqrt{\mu}(y-C)}\quad& \mbox{as}\quad y \to -\infty.
\end{array}\right.
\end{equation}
As an outcome of (2.2) and (2.7), the
solution of (1.1) with initial data $u(x,0)=\varphi (x)$ is
reflectionless at any time $t>0$. Hence the solitary waves of the
Camassa--Holm equation are solitons.

\subsection*{Acknowledgements}

  The author is grateful to A~Constantin for
bringing this problem to his attention and for stimulating
conversations. The author thanks the referee for valuable remarks on
an earlier draft of this note.

\label{lenells-lastpage}

\end{document}